\documentclass[11pt]{article}
\usepackage[utf8]{inputenc}
\usepackage[T1]{fontenc}
\usepackage{graphicx}
\usepackage{longtable}
\usepackage{wrapfig}
\usepackage{rotating}
\usepackage[normalem]{ulem}
\usepackage{amsmath}
\usepackage{amssymb}
\usepackage{capt-of}
\usepackage{hyperref}
\usepackage{xparse}
\usepackage{mathrsfs}
\usepackage{breqn}
\usepackage{braket}

\newcommand{\A}{\mathcal{A}}
\newcommand{\Fg}{\mathcal{F}}

\newcommand{\R}{\mathbb{R}}

\newcommand{\Ric}{\operatorname{Ric}}

\newcommand{\sgn}{\operatorname{sgn}}
\newcommand{\N}{\ensuremath{\mathscr{N}}}
\newcommand{\W}{\ensuremath{\mathscr{W}}}
\newcommand{\Op}{\mathcal{O}}
\newcommand{\we}{\mathop{\scriptstyle\wedge}}

\newcommand*{\de}[1]{\mathop{\mathrm{d}#1}\nolimits}
\newcommand{\dn}[2]{{\mathrm{d}}^{#1}\!{#2}\;}

\NewDocumentCommand\MyAc{ m }{#1}
\DeclareDocumentCommand{\ct}{ t. t, t- t' s s m m m }{
  \RenewDocumentCommand\MyAc{ m }{##1}
  \IfBooleanT{#1}{\RenewDocumentCommand\MyAc{ m }{ \mathring{##1} } }
  \IfBooleanT{#2}{\RenewDocumentCommand\MyAc{ m }{ \tilde{##1} } }
  \IfBooleanT{#3}{\RenewDocumentCommand\MyAc{ m }{ \bar{##1} } }
  \IfBooleanT{#4}{\RenewDocumentCommand\MyAc{ m }{ {##1}' } }
  \IfBooleanTF{#5}
  { \IfBooleanTF{#6} { \hat{\MyAc{\Gamma}}_{{#7}}{}^{\hat{#8}}{}_{\hat{#9}} }{ \hat{\MyAc{\Gamma}}_{{#7}}{}^{{#8}}{}_{{#9}} } }
  { \MyAc{\Gamma}_{{#7}}{}^{{#8}}{}_{{#9}} } }
\DeclareDocumentCommand{\ri}{ t. t, t- s s m m m }{
  \RenewDocumentCommand\MyAc{ m }{##1}
  \IfBooleanT{#1}{\RenewDocumentCommand\MyAc{ m }{ \mathring{##1} } }
  \IfBooleanT{#2}{\RenewDocumentCommand\MyAc{ m }{ \tilde{##1} } }
  \IfBooleanT{#3}{\RenewDocumentCommand\MyAc{ m }{ \bar{##1} } }
  \IfBooleanTF{#4}
  { \IfBooleanTF{#5} { \hat{\MyAc{\mathcal{R}}}_{{#6}}{}^{\hat{#7}}{}_{\hat{#8}} }{ \hat{\MyAc{\mathcal{R}}}_{{#6}}{}^{{#7}}{}_{{#8}} } }
  { \MyAc{\mathcal{R}}_{{#6}}{}^{{#7}}{}_{{#8}} } }
\DeclareDocumentCommand{\bt}{ t. t, t- s s m m m }{
  \RenewDocumentCommand\MyAc{ m }{##1}
  \IfBooleanT{#1}{\RenewDocumentCommand\MyAc{ m }{ \mathring{##1} } }
  \IfBooleanT{#2}{\RenewDocumentCommand\MyAc{ m }{ \tilde{##1} } }
  \IfBooleanT{#3}{\RenewDocumentCommand\MyAc{ m }{ \bar{##1} } }
  \IfBooleanTF{#4}
  { \IfBooleanTF{#5} { \hat{\MyAc{\mathcal{B}}}_{{#6}}{}^{\hat{#7}}{}_{\hat{#8}} }{ \hat{\MyAc{\mathcal{B}}}_{{#6}}{}^{{#7}}{}_{{#8}} } }
  { \MyAc{\mathcal{B}}_{{#6}}{}^{{#7}}{}_{{#8}} } }
\newcommand{\tors}[3]{\mathcal{T}{}_{#1}{}^{#2}{}_{#3}}

\newcommand\UTFSM{Departamento de F\'isica, Universidad T\'{e}cnica Federico Santa Mar\'\i a\\ Casilla 110-V, Valpara\'iso, Chile}
\newcommand\UTFSMmat{Departamento de Matem\'aticas, Universidad T\'{e}cnica Federico Santa Mar\'\i a\\ Casilla 110-V, Valpara\'iso, Chile}

\newcommand{\UdelaR}{Instituto de F\'isica, Facultad de Ciencias\\Igu\'a 4225, esq. Mataojo, 11400 Montevideo, Uruguay.}
\usepackage{cite}
\usepackage[in]{fullpage}
\author{Oscar Castillo-Felisola,\(^{1,*}\) \and Bastian Grez,\(^1\) \and Oscar Orellana,\(^2\) \and Jose Perdiguero,\(^1\) \and Francisca Ramirez,\(^1\) \and Aureliano Skirzewski,\(^3\) \and Alfonso R. Zerwekh.\(^1\)}
\date{\(^1\) \UTFSM \newline \(^2\) \UTFSMmat \newline \(^3\) \UdelaR \newline \(^*\) Email: \href{mailto:o.castillo.felisola@protonmail.com}{o.castillo.felisola@protonmail.com}}
\title{Polynomial affine model of gravity in three-dimensions}
\begin{document}

\maketitle

\begin{abstract}
In this work, we explore a three-dimensional formulation of the
polynomial affine model of gravity, which is a model that extends
general relativity by relaxing the equivalence principle through the
exclusion of the metric from the set of fundamental fields.
In particular, in an attempt to gain insight of the role of the
torsion and nonmetricity in the gravitational models, we consider
homogeneous and isotropic cosmological models, for which their solutions are
classified in a \emph{decisions tree}. We also show a few of these explicit
solutions that allow the definition of (alternative/emergent) metrics derived
from the connection.
\end{abstract}

\section{Introduction}
\label{sec:intro}
During the first quarter of the twentieth century, Einstein built the
general theory of relativity, attempting to render gravitational
interactions and the special theory of relativity compatible. The
resulting model, known as General Relativity, turned to provide a
geometrical interpretation of the gravitational interactions, where
gravitational forces are a manifestation of the curvature of
spacetime. 

General Relativity is so far the most successful model of
gravitational interactions at our disposal, explaining a large amount
of experimental observations \cite{will18_theor_exper_gravit_physic}.
Nonetheless, the completion of General Relativity with the standard
model of particles is insufficient to account for effects such as the
rotation of the Halos of Galaxies
\cite{zwicky37_masses_nebul_clust_nebul,rubin70_rotat_androm_nebul_from_spect,sofue01_rotat_curves_spiral},
and the value of the cosmological constant
\cite{riess98_obser_eviden_from_super_accel,perlmutter99_measur_oemeg_lambd_from_high_redsh_super}.
We should add to that the difficulty to consistently quantise the
model \cite{dewitt67_quant_theor_gravit_i,dewitt67_quant_theor_gravit_ii,dewitt67_quant_theor_gravit_iii,deser74_one_loop_diver_quant_einst_maxwel_field,deser74_nonren_quant_dirac_einst_system,hooft74_one_loop_diver_theor,ashtekar86_new_variab_class_quant_gravit,ashtekar87_new_hamil_formul_gener_relat}.

In order to bypass these vicissitudes, it is necessary either to modify
the matter content of the standard model of particles or to find an
alternative model to General Relativity. In this work, we focus on the
second type of modifications, particularly to enhance the
gravitational sector by allowing a general connection and restricting
the role of the metric.\footnote{In General Relativity the role of the metric is two-folded, it
is the geometric object that defines distances, but also is the field
responsible for the mediation of the gravitational interactions.}

Interestingly, a general notion of (continuous and connected)
manifolds is determined by the \emph{affine connection}, which is a
geometrical object more fundamental than the metric. A manifold
(\(M\)) equipped with an affine connection (\(\ct*{}{}{}\)) is called
an affine manifold, and it admits the notion of parallel transports.
Hence, the Riemannian geometry behind General Relativity is a
particular case of the above when the connection is symmetric in
their lower indices and compatible with a metric tensor field, i.e.,
\(\ct*{[\mu}{\lambda}{\nu]} = 0\) and \(\nabla^{(\hat{\Gamma})} g =
0\). In affinely connected manifolds, it is possible to define
\(\binom{0}{2}\) tensors derived from the connection, e.g., the Ricci
tensor field, for which its symmetric part might serve as a metric field, if
it is non-degenerated \cite{eisenhart27_non_rieman}. Sometimes, these
tensors are referred to as connection descendent or induced or
emergent metric tensor fields.

From a modern perspective, the idea of formulating gravity as an
affine theory is attractive since other fundamental interactions
are gauge theories, for which their fundamental fields are gauge connections,
and such theories are renormalisable. Therefore, an affine formulation
of gravity would bring the theory to the same footing with gauge
theories.

The first proposals of affine gravity were considered by Einstein,
Eddington and Schrödinger 
\cite{einstein23_zur_affin_feldt,einstein23_theor_affin_field,eddington23,schroedinger50_space},
but those models did not offer sufficient phenomenological novelty,
while their treatments were significantly more complex. More
recent affine models of gravity were proposed by Kijowski and
collaborators
\cite{kijowski78_new_variat_princ_gener_relat,ferraris81_gener_relat_is_gauge_type_theor,ferraris82_equiv_relat_theor_gravit,kijowski07_univer_affin_formul_gener_relat},  
Popławski
\cite{poplawski07_nonsy_purel_affin,poplawski07_unified_purel_affin_theor_gravit_elect,poplawski09_gravit_elect,poplawski14_affin_theor_gravit}, 
Krasnov and collaborators
\cite{krasnov07_non_metric_gravit,krasnov08_non_metric_gravit_i,krasnov08_non_metric_gravit_ii,krasnov11_pure_connec_action_princ_gener_relat,delfino15_pure_connec_formal_gravit_lin,delfino15_pure_connec_formal_gravit_feyn}, 
and ourselves
\cite{castillo-felisola15_polyn_model_purel_affin_gravit,castillo-felisola18_einst_gravit_from_polyn_affin_model}.

Albeit the fact that affine formulations of gravity conceptually
differ from general relativity, the predictions obtained from earlier
models do not provide significant differences from those of general
relativity, while their manipulation was harder.\footnote{In the words of Schrödinger, ``For all that I know, no special
solution has yet been found which suggests an application to anything
that might interest us \ldots{}'' \cite{schroedinger50_space}.} It has been
understood that even if two models are equivalent at the dynamical
level, their generalisations might be
inequivalent \cite{krasnov20_formul_gener_relat}. However, subsequent
models predict novel effects.

In the four-dimensional version of the polynomial affine model of
gravity, even if we restrict ourselves to the torsion-free sector of
the model, the vacuum solutions of Polynomial Affine Gravity include
the vacuum solutions of general relativity as a
subset \cite{castillo-felisola18_einst_gravit_from_polyn_affin_model}.
Moreover, some vacuum solutions of our affine model account for
effects that in general relativity are induced by the presence of
matter. This is interpreted as a mimicking of matter effects by the
non-Riemannian structure of 
geometry \cite{castillo-felisola18_cosmol}. Although the underlying
geometry of the polynomial affine model of gravity does not require
the existence of a metric ab\textasciitilde{}initio, it is possible to define
connection-descendent metrics,\footnote{We also refer to these metrics as \emph{emergent} metrics.} allowing to distinguish between null
and non-null \emph{geodesics} (self-parallel), which would describe the
free-fall trajectories for mass-less and massive
particles \cite{castillo-felisola20_emerg_metric_geodes_analy_cosmol}.
What is worth noticing is the existence of a connection-descendent metric
permits  making contact with other quantities of interest in
cosmological and astronomical/astrophysical applications, such as the
red shift.

Moved by the moral from the early attempts to build affine models of
gravity, it is convenient to analyse the model in lower dimensions,
where the number of parameter is in general diminished. These models
are a playground to test methods that would be applied in four
dimensions, but also might be applied in other branches of physics
since some gauge theories in three dimensions can be interpreted as
the high-temperature limit of four-dimensional models
\cite{weinberg76_under}.

Historically, the three-dimensional version of General Relativity was
firstly studied by Staruszkiewicz
\cite{staruszkiewicz63_gravit_theor_three_dimen_space}, who considered
Schwarzschild-like solutions and noticed a relation between the
presence of massive point particles and conical singularities. It was
soon understood that in three-dimensional General Relativity there
were no propagating degrees of freedom as a consequence of a relation
between the Riemann and Ricci tensors (they are not independent since
the conformal Weyl tensor in three dimensions is trivial). Due to
this, the interest for three-dimensional gravity decreased. It was
only after a series of papers by Deser, Jackiw and Templeton
\cite{deser82_three_dimen_massiv_gauge_theor,deser82_topol_massiv_gauge_theor,deser84_three_dimen_cosmol_gravit,deser84_three_dimen_einst_gravit},
who added a nontrivial topological term to the action, that
three-dimensional gravity re-emerged. In those works, the authors
showed that the additional term induced a massive model of gravity
dubbed Deser--Jackiw--Templeton gravity (or DJT for short).

Later, it was shown that the three-dimensional gravity modified by the
Pontryagin--Chern--Simons Lagrangian is equivalent to a Yang--Mills
theory (called Chern--Simons gravity), and its perturbative expansion
was renormalisable
\cite{witten88_dimen_gravit_as_exact_solub_system}.\footnote{An update on the original ideas in this paper can be found in
Ref. \cite{witten07_three_dimen_gravit_revis}.} Later, Bañados,
Teitelboim and Zanelli found the first black hole solution
\cite{banados92_black_hole_three_dimen_time}, disproving the \emph{triviality}
of the classical three-dimensional gravity. After these there was a
boost in the search of exact solutions
\cite{garcia-diaz17_exact_solut_three_dimen_gravit} and analysis of
their quantum aspects \cite{carlip98_quant}.

From all of the above, it is possible to conclude that the
non-triviality of three-dimensional gravity comes from topological
aspects. However, when one considers affinely connected manifolds, it
is possible to define the \emph{projective Weyl tensor field}:
\begin{equation*}
  W_{\mu\nu}{}^\lambda{}_\rho = \ri{\mu\nu}{\lambda}{\rho} -
  \frac{1}{D-1} \left( \ri{\nu\rho}{}{} \delta^\lambda_\mu -
    \ri{\mu\rho}{}{} \delta^\lambda_\nu \right),
\end{equation*}
without the use of a metric tensor. The projective Weyl tensor, unlike
the conformal Weyl tensor (which requires a metric and has terms,
with the scalar curvature), does not vanish necessarily. Therefore,
gravitational models involving the affine connection might 
express dynamical effects unaccounted for by General Relativity. An example
of the modifications to the dynamics comes from the existence of
nonmetricity which, unlike the torsion, introduces a term that is
responsible of the non-coincidence of geodesics and self-parallel
curves. Moreover, such a quantity cannot be set to zero through a
coordinate transformation due to its tensor character, i.e., it is only 
zero in a frame if it vanishes in any other frame.

The aim of this study is to review some of the aspects of the
polynomial affine model of gravity introduced in Ref.
\cite{castillo-felisola15_polyn_model_purel_affin_gravit}, focusing our
attention on the three-dimensional scenario. In this context, we
find ansätze compatible with the cosmological principle, which allows us
to solve the field equations. Interestingly, some of the solutions
admit non-degenerated derived metrics.

\section{What is the polynomial affine model of gravity?}
\label{sec:what_is_pag}
Our purpose is to build an alternative model of gravity where the
degrees of freedom mediating the interaction come from a fundamental
affine connection, instead of coming from the metric tensor field. In
order to achieve our goals, we suppressed the use of the metric in the
formulation of the action and focused on terms that are polynomials
in the fields (and their derivatives). As a consequence, the number of 
possible terms involved in the action is finite, as we shall show by
an analysis of the index structure later in this section.

Starting from the affine connection, \(\ct*{\mu}{\lambda}{\rho}\), we
can decompose it into its irreducible components by separating the
symmetric and skew-symmetric parts (in the lower indices),
\begin{equation*}
  \ct*{\mu}{\lambda}{\nu} 
  =
  \ct*{(\mu}{\lambda}{\nu)} + \ct*{[\mu}{\lambda}{\nu]}
  =
  \ct{\mu}{\lambda}{\nu} + \bt{\mu}{\lambda}{\nu} + \A_{[\mu}\delta^\lambda_{\nu]}.
  \label{eq:conn_decomp}
\end{equation*}
In the last equality, we split the skew-symmetric component of the
connection further into its trace, \(\A\), and its traceless part,
\(\bt{}{}{}\). Hereafter, we denote the symmetric component of the
connection by \(\ct{}{}{}\) (without the hat). It is worth mentioning
that the skew-symmetric part of the connection is related to the
torsion tensor field; therefore, both \(\A\) and \(\bt{}{}{}\) are
the building blocks of torsion.

Our guideline to build the action functional would be the invariance
under diffeomorphisms, and the field content of the model would be the
symmetric connection (\(\Gamma\)), the trace of the torsion (\(\A\)),
and the traceless torsion (\(\bt{}{}{}\)). Note that the
torsion-descendent fields are tensor fields, while \(\Gamma\) is
(still) a connection. Consequently, the symmetric connection cannot
(in principle) enter directly to the action, but instead it should
enter through the covariant derivative, \(\nabla\).

In order to build an action functional successfully, we require a
volume form in the affine manifold. Such volume form is a
generalisation of the well-known Riemannian volume form, which is
defined by the metric tensor field, \(\de{V}_g = \dn{3}{x} \sqrt{g}\), 
where the symbol \(g\) represents the (absolute value of the)
determinant of the metric tensor field. We denote the affine volume by
\(\de{V}^{\mu \nu \lambda} = \de{x}^\mu \we \de{x}^\nu \we
\de{x}^\lambda\).

How do we build the action with these ingredients? Since the action is
the integral of the Lagrangian, which is a scalar density, we should
write down all possible terms that are polynomial in the fields, which
can be contracted to form scalar densities. Since \(\A\),
\(\bt{}{}{}\), and \(\nabla\) have tensor character and the volume
form is the only element with \emph{weight}, we can classify the terms in
the action by an analysis of the indices structure.\footnote{In previous articles we refer to this as \emph{dimensional analysis}.}

Note that all ``tensor'' fields have a simple net lower index
(e.g., the \(\bt{}{}{}\) field has one upper and two lower indices; thus,
it would contribute to a net single index), while the volume form has
three upper indices. Let us define the \emph{index balance operator} as \(\N\),
which returns the value \(+1\) for each upper index and \(-1\) for
lower indices, and the \emph{weight operator} as \(\W\), which returns the
weight of the tensor density. Then, an operator of the following form
is obtained:
\begin{equation*}
  \Op = \A^m \bt{}{}{}^n \nabla^p \de{V}^q,
\end{equation*}
where the superindices denote the power of the element (number of
times it might appear), and it can be part of the Lagrangian when the net
index balance is zero and its weight is one; thus, thus it should satisfy
the following:
\begin{equation*}
  \N(\Op) = 3q - m - n - p = 0, \qquad \W(\Op) = q = 1,
\end{equation*}
if the operator could take part in the action.

Now that we know the restrictions in the power of the fields, we write
down all possible contractions, and use the symmetries of the
fields to eliminate redundant terms. The most general action (up to
boundary terms) is given by the following (see Ref.
\cite{castillo-felisola15_polyn_model_purel_affin_gravit}):
\begin{dmath}
  \label{eq:action}
  S = \int \de{V}^{\alpha\beta\gamma} \bigg(
  B_1 \, \A_\alpha \A_\mu \bt{\beta}{\mu}{\gamma}
  + B_2 \, \A_\alpha \Fg_{\beta\gamma}
  + B_3 \, \A_\alpha \nabla_\mu \bt{\beta}{\mu}{\gamma}
  + B_4 \, \bt{\alpha}{\mu}{\nu} \bt{\beta}{\nu}{\lambda} \bt{\gamma}{\lambda}{\mu}
  + B_5 \ri{\alpha\beta}{\mu}{\mu} \A_\gamma
  + B_6 \, \ri{\mu\alpha}{\mu}{\nu} \bt{\beta}{\nu}{\gamma}
  + B_7 \, \ct{\alpha}{\mu}{\mu} \partial_\beta \ct{\gamma}{\nu}{\nu}
  + B_8 \, \left( \ct{\alpha}{\mu}{\nu} \partial_\beta \ct{\gamma}{\nu}{\mu} +
    \frac{2}{3} \ct{\alpha}{\mu}{\nu} \ct{\beta}{\nu}{\lambda} \ct{\gamma}{\lambda}{\mu} \right)
  \bigg),
\end{dmath}
where the topological Chern--Simons terms have been included
explicitly. In the above, we use the symbols \(\Fg\) to denote the \emph{field
strength} of the \(\A\)-field, i.e., \(\Fg_{\beta\gamma} =
\partial_{\beta} \A_\gamma - \partial_\gamma \A_\beta\), and we use
\(\ri{}{}{}=\ri{}{\Gamma}{}\) to denote the curvature of the symmetric
connection.

In comparison with the action of General Relativity, the Eq.
\eqref{eq:action} seems unpleasant at first. However, we should have in
mind that the Hilbert--Einstein action is far from being the \emph{most
general} one compatible with the symmetries. For this reason it is
useful to highlight some nice features of the polynomial affine model
of gravity, which can be read from its action: (i) Similarly for the other
fundamental interactions, the fundamental field is a connection; (ii)
All coupling constants are dimensionless, which is desirable from
the point of view of Quantum Field Theory, since the superficial
degree of divergence vanishes; (iii) At least at a classical level,
the model seems to exhibit scale invariance; (iv) The number of
possible terms in the action is finite (we usually refer to this
property as the \emph{rigidity} of the model), giving the impression
that in the hypothetical scenario of quantisation, the model would be
renormalisable, since all counter-terms should have the form of
terms already present in the original action.

The field equations derived from the action in Eq. \eqref{eq:action} are
as follows:
\begin{dmath}
  2 B_1 \A_\alpha \bt{\nu}{\alpha}{\rho} + 2 B_2
  \Fg_{\nu\rho} + B_3 \nabla_\mu \bt{\nu}{\mu}{\rho} + B_5
  \ri{\nu\rho}{\mu}{\mu} = 0,
  \label{eq:feq_A}
\end{dmath}
\begin{dmath}
  2 B_1 \A_\nu \A_\rho - 2 B_3 \nabla_{(\nu} \A_{\rho)} + 3 B_4
  \bt{\nu}{\mu}{\sigma} \bt{\rho}{\sigma}{\mu} + 2 B_6
  \ri{\mu(\nu}{\mu}{\rho)} = 0,
  \label{eq:feq_B}
\end{dmath}
\begin{dmath}
  B_3 \A_\mu \bt{\rho}{\nu}{\sigma}
+ B_5 \left( \delta_\mu^\nu
    \Fg_{\rho\sigma} +  \delta_{[\rho}^\nu \Fg_{\sigma]\mu} \right)
+ B_6 \left( 2 \delta_{\mu}^{\nu} \nabla_{\tau}{ \bt{\rho}{\tau}{\sigma} }
    + \delta_{\rho}^{\nu} \nabla_{\tau}{ \bt{\sigma}{\tau}{\mu} }
    + \delta_{\sigma}^{\nu} \nabla_{\tau}{ \bt{\mu}{\tau}{\rho} }
    \right)
+ B_7 \left( \delta_\mu^\nu
    \ri{\rho\sigma}{\lambda}{\lambda} + \delta_{[\rho}^\nu 
    \ri{\sigma]\mu}{\lambda}{\lambda} \right)
+ B_ 8 \left( \ri{\rho\sigma}{\nu}{\mu} + \delta_{[\rho}^\nu
    \ri{\sigma]\lambda}{\lambda}{\mu} \right) = 0.
  \label{eq:feq_G}
\end{dmath}

We have to mention that the field equations presented above are
obtained by a naive optimization of the action functional,
without taking into account the possible existence of second class
constraints that could result in the necessity of adding boundary
contributions to the action.

In order to solve the field equations, we use the same strategy as in
General Relativity to provide an ansatz that must be compatible
with the requirements of our problem. 

\section{Building the ansätze}
\label{sec:ansatz}
Riemannian geometry is the arena where General Relativity lays its
grounds, and the \emph{fundamental} geometrical object in Riemannian
geometry is the metric tensor field. Since the metric tensor field is
related to the notion of distance, in the easiest examples it is
straightforward to propose ansätze. However, for less trivial
examples one utilises the notion of the Lie derivative and its relation
with the concept of symmetries.

Although the action of the Lie derivative on tensors (and even tensor
densities) is commonly known \cite{wald10_gener_relat}, it is less
known that it is possible to define the Lie derivative of an affine
connection
\cite{schouten13_ricci,lovelock89_tensor,ortin15_gravit_strin}, given
by the following relation ,
\begin{dmath}
  \pounds_V \ct*{\mu}{\lambda}{\nu} =
  V^\sigma \partial_\sigma \ct*{\mu}{\lambda}{\nu}
  - \ct*{\mu}{\sigma}{\nu} \partial_\sigma V^\lambda
  + \ct*{\sigma}{\lambda}{\nu} \partial_\mu V^\sigma
  + \ct*{\mu}{\lambda}{\sigma} \partial_\nu V^\sigma
  + \frac{\partial^2 V^{\lambda}}{\partial x^{\mu} \partial x^\nu }
  = \hat{\nabla}_\mu \hat{\nabla}_\nu V^\lambda
  + \ri*{\rho\mu}{\lambda}{\nu} V^\rho
  - 2 \hat{\nabla}_\mu \left( \ct{[\nu}{\lambda}{\rho]} V^\rho  \right).
  \label{eq:lie_der_connection}
\end{dmath}
Above, the vector field \(V\) represents a generator of the symmetry
group, i.e., in mathematical terminology, \(V\) defines the symmetry
flow. Note that the second line is a covariant expression, since the
skew-symmetric part of the connection (i.e., the torsion) is a tensor
field.

Therefore, the strategy to find ansätze would be the following: (i)
Define the desired symmetries in concordance with the physical
problem we would like to solve; (ii) Obtain the vector fields
associated to the generators of the symmetry group and express them
in the coordinate system of preference; (iii) Restrict the
coefficients of the affine connection by solving the set of equations
defined by \(\pounds_V \ct*{\nu}{\lambda}{\rho} = 0\). Note that this
represents a set of \(27\) partial differential equations for each
vector \(V\). 

In order to illustrate the procedure, we use \emph{spherical} coordinates,
\((t,r,\varphi)\), and show that the general affine connection is
compatible with isotropy and the cosmological principle (i.e.,
isotropic and homogeneous).

In three-dimensions, the vector fields generating the rotation and
translations along a two-dimensional subspace of constant curvature
are represented by the following:
\begin{equation}
  \begin{aligned}
    J & = 
    \begin{pmatrix}
      0 & 0 & 1
    \end{pmatrix},
    \\
    X & = 
    \sqrt{1 - \kappa r^2} 
    \begin{pmatrix}
      0 & \cos \varphi & - \frac{1}{r} \sin \varphi
    \end{pmatrix},
    \\
    Y & = 
    \sqrt{1 - \kappa r^2} 
    \begin{pmatrix}
      0 & \sin \varphi &  \frac{1}{r} \cos \varphi
    \end{pmatrix},
  \end{aligned}
  \label{eq:Killing_vectors}
\end{equation}
where \(\kappa = -1, 0, +1\) is the normalised Gaußian curvature of
the two-dimensional subspace.

\subsection{Isotropic connection}
\label{sec:ansatz_isotropic}
Straightforwardly, one can observe from Eq. \eqref{eq:lie_der_connection}
that, since the vector field \(J\) is constant, the non-homogeneous
contribution to the Lie derivative vanishes; thus, it coincides
with the Lie derivative of a \(\binom{1}{2}\)-tensor.

Moreover, the explicit relations derived from the condition of the
vanishing Lie derivative of the connection along the vector \(J\) are
as follows:
\begin{equation}
  \pounds_J \ct*{\mu}{\lambda}{\nu} = \partial_\varphi \ct*{\mu}{\lambda}{\nu} = 0.
  \label{eq:isotropic_connection}
\end{equation}
The solution is that none of the components of the affine connection
depend on the angular coordinate \(\varphi\). Unfortunately, this
symmetry argument does not kill any of the components of the
connection! Consequently, using only an argument of isotropy, it would
be almost impossible to find an analogous case of the Schwarzschild black
hole---which in three-dimensions it is called
Bañados--Teitelboim--Zanelli (BTZ) black hole
\cite{banados92_black_hole_three_dimen_time,carlip05_quant_gravit_dimen,zanelli16_chern,garcia-diaz17_exact_solut_three_dimen_gravit}. 

\subsection{Isotropic and homogeneous connection}
\label{sec:ansatz_cosmo}
In order to find the affine connection compatible with the
cosmological principle, we have to solve the conditions of the vanishing
Lie derivative of the connection along the vector fields \(X\) and
\(Y\). This procedure is long and tedious, but straightforward, and we
refer to the interested reader to check Refs.
\cite{castillo-felisola18_beyond_einstein,castillo-felisola18_cosmol}
to see more details.

After solving the conditions, the irreducible components of the affine
connection are given by the following:
\begin{equation}
  \begin{aligned}
    \ct{t}{t}{t} & = j(t),
    &
    \ct{i}{t}{j} & = g(t) S_{ij},
    \\
    \ct{i}{k}{j} & = \gamma_i{}^k{}_j,
    & \ct{t}{i}{j} = \ct{j}{i}{t} & = h(t) \delta^i_j + f(t) S^{ik} \epsilon_{kj} \frac{r}{\sqrt{1 - \kappa r^2}}, 
  \end{aligned} 
  \label{eq:homotropic_connection} 
\end{equation}
where \(f\), \(g\), \(h\), and \(j\) are functions of time, while
\(S_{ij}\) and \(\gamma_{i}{}^{j}{}_{k}\) are the two-dimensional rank
two symmetric tensor and connection compatible with isotropy and
homogeneity, defined by the following:
\begin{equation*}
  S_{ij} =
  \begin{pmatrix} 
    \frac{1}{1 - \kappa r^2} & 0 \\
    0 & r^2
  \end{pmatrix},
\end{equation*}
and 
\begin{align*}
  \gamma_{r}{}^{r}{}_{r} & = \frac{\kappa r}{1 - \kappa r^2},
  &
  \gamma_{\varphi}{}^{r}{}_{\varphi} & = - r ( 1- \kappa r^2 ),
  &
  \gamma_{r}{}^{\varphi}{}_{\varphi} & = \frac{1}{r},
  &
    \gamma_{\varphi}{}^{\varphi}{}_{r} & = \frac{1}{r}.
\end{align*}
If one compares with the Levi-Civita connection obtained from a
Friedman--Robertson--Walker metric, one notices that the function
\(j\) is zero, the function \(f\) is something entirely new (and it
exists solely in three-dimensions), and the functions \(g\) and \(h\)
are defined in terms of the scale factor \(a\), as \(g = a \dot{a}\)
and \(h = \dot{a} / a\).

It is possible to show that in the affine case, a reparametrisation of
the time allows us to set the function \(j = 0\)
\cite{castillo-felisola21_aspec_polyn_affin_model_gravit_three}.

The nonvanishing components of the \(\bt{}{}{}\)-field are as follows:
\begin{equation}
  \begin{aligned}
    \bt{\varphi}{t}{r} = - \bt{r}{t}{\varphi} & = \xi(t) \frac{r}{\sqrt{1 - \kappa r^2}},
    \\
    \bt{t}{r}{\varphi} = - \bt{\varphi}{r}{t} & = \psi(t) r \sqrt{1 - \kappa r^2},
    \\
    \bt{r}{\varphi}{t} = - \bt{t}{\varphi}{r} & = \frac{\psi(t)}{r \sqrt{1 - \kappa r^2}},
  \end{aligned}
  \label{eq:homotropic_bt}
\end{equation}
while the nonvanishing component of the \(\A\)-field is \(\A_t =
\eta(t)\). 

\section{Cosmological solutions}
\label{sec:cosmo_solutions}
Now that we have the general form of the fields, we substitute such
ansatz into the field equations in Eq. \eqref{eq:feq_A}, \eqref{eq:feq_B}
and \eqref{eq:feq_G}. The resulting system of equations is
\begin{dgroup}[noalign]
  \begin{dmath}
    2 B_8 g f - B_3 \xi \eta = 0,
    \label{eq:oo1}
  \end{dmath}
  \begin{dmath}
    B_6 (2 g \psi + \dot{\xi}) - B_8 f g = 0,
    \label{eq:oo2}
  \end{dmath}
  \begin{dmath}
    B_8 (2 g h + \kappa - \dot{g}) = 0,
    \label{eq:oo3}
  \end{dmath}
  \begin{dmath}
    B_3 \eta \psi + B_8 (2 h f + \dot{f}) = 0,
    \label{eq:oo4}
  \end{dmath}
  \begin{dmath}
    B_1 \eta^2 - B_3 \dot{\eta} + 3 B_4 \psi^2 - 2 B_6 (\dot{h} + h^2
    - f^2) = 0,
    \label{eq:oo5}
  \end{dmath}
  \begin{dmath}
    B_3 g \eta - 3 B_4 \psi \xi + B_6 (\kappa + \dot{g}) = 0,
    \label{eq:oo6}
  \end{dmath}
  \begin{dmath}
    2 B_1 \eta \xi + B_3 (2 g \psi + \dot{\xi}) = 0.
    \label{eq:oo7}
  \end{dmath}
  \label{eq:cosmological_feqs}
\end{dgroup}
Notably, the Eq. \eqref{eq:oo1} is algebraic. However, we can obtain
other algebraic relations from Eqs. \eqref{eq:oo2} and \eqref{eq:oo7}, and
from Eqs. \eqref{eq:oo3} and \eqref{eq:oo6}. These expressions are
\begin{dgroup}[noalign]
  \begin{dmath}
    2 B_8 g f - B_3 \xi \eta = 0,
    \label{eq:ooa}
  \end{dmath}
  \begin{dmath}
    \frac{B_8}{B_6} g f + 2 \frac{B_1}{B_3} \xi \eta = 0,
    \label{eq:oob}
  \end{dmath}
  \begin{dmath}
    3 \frac{B_4}{B_6} \psi \xi - 2 g h - \frac{B_3}{B_6} g \eta = 2
    \kappa.
    \label{eq:ooc}
  \end{dmath}
  \label{eq:oo_eqs}
\end{dgroup}
Equations \eqref{eq:ooa} and \eqref{eq:oob} can be seen as a system of
equations for variables \(f\) and \(\eta\) as functions of \(g\)
and \(\xi\), but their independence is dictated by the determinant of
the following system coefficients,
\begin{equation}
  \Omega = B_8 \left[ 4 \frac{B_1}{B_3} + \frac{B_3}{B_6} \right] g \xi.
  \label{eq:oo_omega}
\end{equation}

A thorough search for solutions requires taking into account all
possible cases, which can be represented by \emph{decision trees}. We shall
sketch the types of solutions in two cases: (i) when the determinant
in Eq. \eqref{eq:oo_omega} is nonvanishing, which requires \(f = \eta
=0\); and (ii) when the determinant \(\Omega\) vanishes, for which the
decision tree has many branches.

\begin{center}
\includegraphics[width=.6\textwidth]{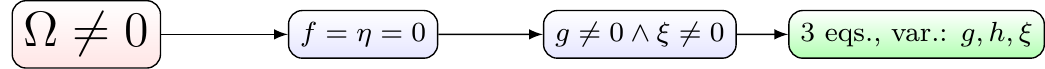}
\end{center}

\begin{center}
\includegraphics[width=.9\textwidth]{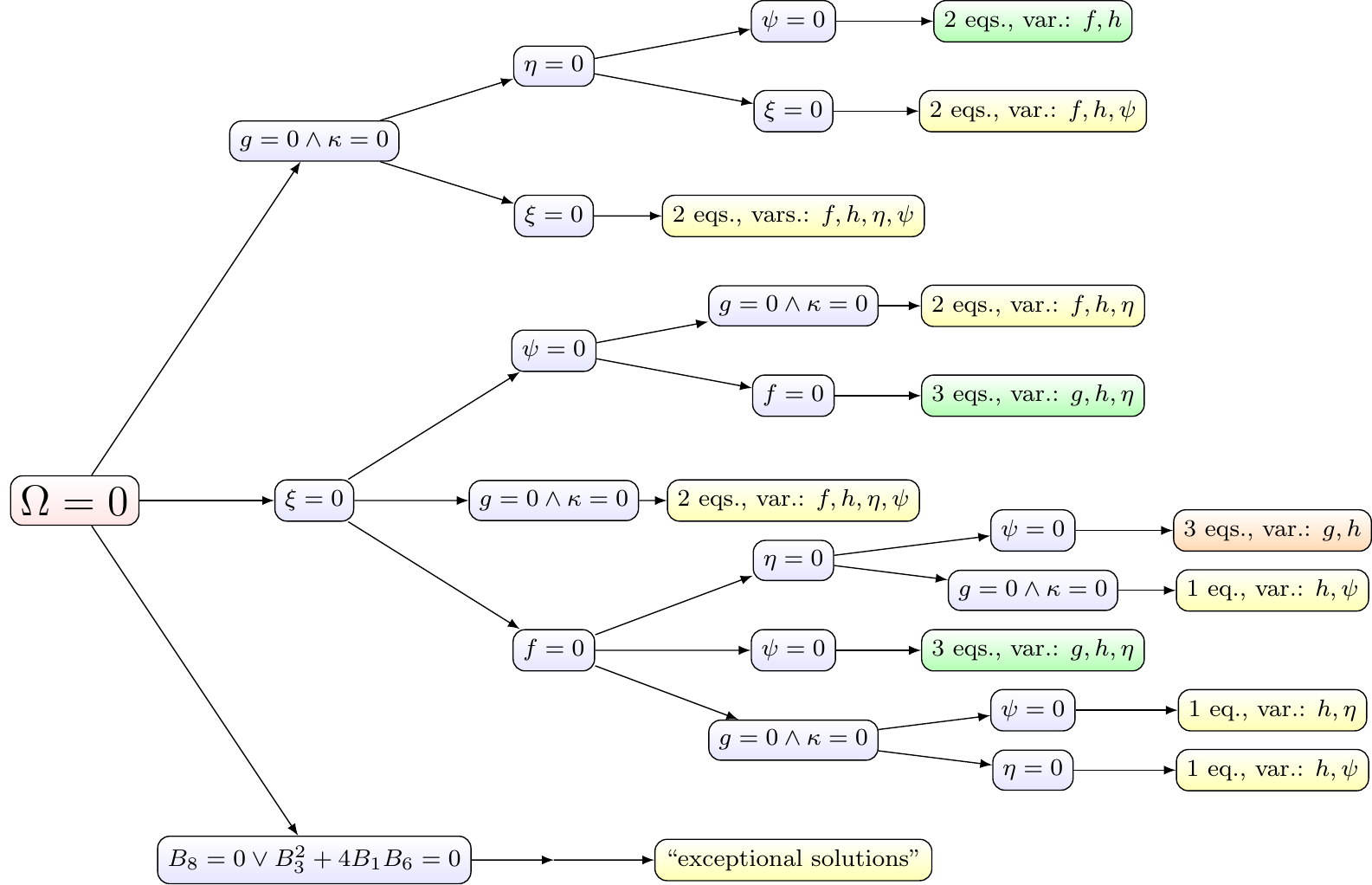}
\end{center}

Interestingly, from the solutions obtained in the scan, just a few of
them have all the fields turned on
\cite{castillo-felisola21_aspec_polyn_affin_model_gravit_three}. Below we present some of these
\emph{exceptional} solutions.

Firstly, the case with \(g = \kappa = \xi = 0\) has as nontrivial
field equations the set
\begin{dgroup}
  \begin{dmath}
    \dot{f} + 2 f h = - \frac{B_3}{B_8} \eta \psi,
    \label{eq:f_param_eta_psi}
  \end{dmath}
  \begin{dmath}
    \dot{h} + h^2 - f^2 = \frac{1}{2 B_6} \left( B_1 \eta^2 - B_3
      \dot{\eta} + 3 B_4 \psi^2 \right),
    \label{eq:h_param_eta_psi}
  \end{dmath}
  \label{eq:fh_param_eta_psi}
\end{dgroup}
which are solved when all the functions are inversely proportional to
\(t\). Note that with this configuration, \(\Gamma\), \(\A\), and
\(\bt{}{}{}\) are nonvanishing.

Another branch of solutions is found when the coupling constants are
not all independent, e.g., for \(B_3^2 + 4 B_1 B_6 = 0\), condition
\(\Omega = 0\) is satisfied without fixing \(g\) or \(\xi\). Under
these conditions, we observed a couple of exceptional solutions.

First, for \(B_1 = B_3 = 0\), the solution is characterised by the
functions 
\begin{equation}
  \begin{aligned}
    f(t) & = 0,
    &
    g(t) & = - \kappa t + C_g,
    &
    h(t) & = \frac{1}{t - \kappa C_g},
    \\
    \xi(t) & = \sqrt{- \frac{2 B_6}{3 B_4}} C_g,
    &
    \psi(t) & = 0,
    &
    \eta(t) & = \text{arbitrary},
  \end{aligned}
  \label{eq:exceptional_sol1_kn0}
\end{equation}
for \(\kappa \neq 0\). While for \(\kappa = 0\), the functions
defining the connection are as follows:
\begin{equation}
  \begin{aligned}
    f(t) & = 0,
    &
    g(t) & = C_m t + C_g,
    &
    h(t) & = \frac{C_m}{2 (C_m t + C_g)},
    \\
    \xi(t) & = \sqrt{- \frac{2 B_6}{3 B_4}} (C_m t + C_g),
    &
    \psi(t) & = \sqrt{- \frac{B_6}{6 B_4}} \frac{C_m}{C_m t + C_g},
    &
    \eta(t) & = \text{arbitrary}.
  \end{aligned}
  \label{eq:exceptional_sol1_k0}
\end{equation}

Secondly, for \(\kappa = 0\) we proposed the ansatz \(g(t) = t^n\)
with \(n \in \R - \Set{-2, \left[\tfrac{1-\sqrt{33}}{4},
\tfrac{1+\sqrt{33}}{4}\right]}\), with the condition
\begin{equation*}
  B_4 = - \frac{8 B_6^3 (n+2)}{3 B_8^2 \left(2 n^3-3 n^2-3 n+4\right)}.
\end{equation*}

The functions defining the connection are as follows,
\begin{equation}
  \label{eq:exceptional_sol2_k0}
  \begin{aligned}
    f(t) & = \frac{\sqrt{2 n^2 - n - 4} \sgn(B_8)}{2 t},
    &
    g(t) & = t^n,
    &
    h(t) & = \frac{n}{2 t},
    \\
    \xi(t) & = \frac{\sqrt{2 n^2 - n - 4} \sgn(B_8) t^n}{2 B_6},
    &
    \psi(t) & = \frac{(n - 1) \sqrt{2 n^2 - n - 4} \sgn(B_8)}{4 t B_6},
    &
    \eta(t) & = \frac{2 B_6}{B_3 t}.
  \end{aligned}
\end{equation}

The solutions shown above are interesting because they admit some sort
of connection-descendent metric. In particular, the solution from Eq.
\eqref{eq:exceptional_sol2_k0} admits several \emph{emergent metrics}, for
example, (i) The Ricci tensor field, and (ii) The Popławski metric
\cite{poplawski14_affin_theor_gravit},\footnote{We shall denote the Popławski metric by \(\operatorname{Popl}\)
or \(g^{(P)}\).} defined from the torsion
tensor field as \(g_{\mu\nu} = \tors{\mu}{\lambda}{\rho}
\tors{\nu}{\rho}{\lambda}\), which are in general non-degenerated.
Without further details, for this solution the metrics are of the
following form:
\begin{equation*}
  \Ric = - \frac{A(n)}{t^2} \de{t} \otimes \de{t}
      + n t^{n-1} \left( \frac{\de{r} \otimes \de{r}}{1 - \kappa r^2} + r^2 \de{\varphi} \otimes \de{\varphi} \right),
\end{equation*}
and
\begin{equation*}
  \operatorname{Popl} = - \frac{F(n,B_i)}{t^2} \de{t} \otimes \de{t}
      + {G(n,B_i) t^{n-1}} \left( \frac{\de{r} \otimes \de{r}}{1 - \kappa r^2} + r^2 \de{\varphi} \otimes \de{\varphi} \right),
\end{equation*}
where \(A\), \(F\) and \(G\) are numerical factors depending on
coefficients \(B_i\) and the value of \(n\)
\cite{castillo-felisola21_aspec_polyn_affin_model_gravit_three}. 

\section{Discussion and concluding remarks}
\label{sec:discussion}
In this article, we have briefly reviewed the polynomial affine model
of gravity, which attempts to model the gravitational interaction as
mediated by the affine connection. Its action functional is built up
without the use of a metric tensor field by explicitly excluding it
from the action.

Despite the fact that an affine model of gravity brings gravitation a
step closer to the formulation of gauge theories, it is interesting
that the number of possible terms in the action [see Eq.
\eqref{eq:action}] is finite.\footnote{Note that once the metric is included, as in General
Relativity, the number of possible terms is naturally unbound, since
we allow the contraction of the indices that are initially on the same
footing. As an example of these terms we have the scalar curvature,
\(\ri{}{}{}\), and its powers, \(\ri{}{n}{}\), opening the window to
the \(f(\ri{}{}{})\)-models.}

Remarkably, even if we start with an affinely connected manifold,
without referring to a fundamental metric, there are symmetric
\(\binom{0}{2}\) tensors derived from the connection that might
play the role of metric tensor fields when non-degenerated.

The first connection-descendent object that might be a metric is the
symmetric Ricci tensor field \cite{eisenhart27_non_rieman}. In fact,
Einstein and Eddington proposed independently their affine model of
gravity by using the square-root of the determinant of the Ricci, i.e.,
\(\sqrt{\Ric}\), as Lagrangian density. Another emergent metric is
given by a quadratic combination of the torsion tensor,
\(g^{(P)}_{\mu\nu} = \tors{\mu}{\lambda}{\rho}
\tors{\nu}{\rho}{\lambda}\), which was considered by Popławski in Ref.
\cite{poplawski14_affin_theor_gravit}. We can define a Popławski-like
metric by using the \(\bt{}{}{}\)-field, as \(g^{(\bt{}{}{})}_{\mu\nu} =
\bt{\mu}{\lambda}{\rho} \bt{\nu}{\rho}{\lambda}\).

In addition, in three-dimensions, the dual of the \(\bt{}{}{}\)-field,
\(\mathfrak{B}^{\lambda\rho} = \frac{1}{2} \epsilon^{\mu\nu\lambda}
\bt{\mu}{\rho}{\nu}\), is a symmetric \(\binom{2}{0}\)-tensor density.
When this tensor density is non-degenerated, it might be interpreted as
an analogous case of the Riemannian \(\sqrt{g} g^{\mu\nu}\). Hence, such
quantity induces a notion of distance, i.e., introduces a metric.

The general field equations are a clear generalisation of the Einstein
equations; in particular, the field equation for the
\(\bt{}{}{}\)-field [see Eq. \eqref{eq:feq_B}] is the analogous case for
the standard equations in General Relativity, where the following term
is provided:
\begin{equation*}
  2 B_1 \A_\nu \A_\rho - 2 B_3 \nabla_{(\nu} \A_{\rho)} + 3 B_4 \bt{\nu}{\mu}{\sigma} \bt{\rho}{\sigma}{\mu}
\end{equation*}
and it might be interpreted as the energy-momentum tensor field of a
geometrically induced matter content. It is worth mentioning that the
Ricci tensor field in Eq. \eqref{eq:feq_B} in general encodes
information about nonmetricity, since the symmetric connection,
\(\ct{}{}{}\), is not necessarily compatible with emergent
metrics
\cite{castillo-felisola20_emerg_metric_geodes_analy_cosmol,castillo-felisola21_aspec_polyn_affin_model_gravit_three}.

In the sense of the last paragraph, the polynomial affine model of
gravity provides nontrivial solutions in the absence of matter, i.e.,
there are matter-like effects induced by non-Riemannian geometrical
quantities.

Lastly, although\textasciitilde{}the Hamiltonian analysis of the polynomial affine
model of gravity has not been completed, we have reasons to think that
in this model there are propagating degrees of freedom, unlike the
three-dimensional version of general\textasciitilde{}relativity.

Before finishing, we would like to remark some of the features that,
in\textasciitilde{}four dimensions, differentiate the polynomial affine model of
gravity from general\textasciitilde{}relativity.

We know that the four-dimensional version of Polynomial Affine Gravity
yields to gravitational interactions (forces) and vacuum solutions
that differ from those of general relativity. Nevertheless, our vacuum
solutions contain effects that in general relativity can be obtained
solely in presence of matter. Ergo, the\textasciitilde{}previously mentioned
non-Riemannian matter-like effects might (at least partially) account
in the four-dimensional context, for\textasciitilde{}the dark sector of the\textasciitilde{}Universe.

\subsubsection*{Acknowledgements}
\label{sec:org49ad176}
The work OCF and ARZ is sponsored by the ``Centro Cient\'ifico y
Tecnol\'ogico de Valpara\'iso'' \mbox{(CCTVal)}, funded by the Chilean
Government through the Centers of Excellence Base Financing Program of
Agencia Nacional de Investigaci\'on y Desarrollo (ANID), by grant ANID
PIA/APOYO AFB180002. This work was funded by ANID Millennium Science
Initiative Program ICN2019\_044, and benefited from grant
PI\_LI\_19\_02 from the Universidad T\'ecnica Federico Santa Mar\'ia.
The~work of FR is funded by ANID BECAS/DOCTORADO NACIONAL 21211990. BG
acknowledges financial support from UTFSM master scholarship No.
034/2021.

We gratefully acknowledge the constructive comments of  anonymous
referees, which helped to improve the quality of the final version of
the manuscript.

\bibliographystyle{unsrt}
\bibliography{References} 
\end{document}